\documentstyle[11pt,fleqn]{article}
\def\beeq{\begin{equation}}
\def\eneq{\end{equation}}
\def\beeqa{\begin{eqnarray}}
\def\eneqa{\end{eqnarray}}

\def\soc{{\rm C}_{60}}

\def\eo{E_{\rm O}}

\def\eps{\varepsilon}

\addtocounter{section}{-1}
\begin{document}

\begin{center}

{\large{\bf{C$_{60}$ with O defects:\\ Localized states in the gap and
oxygen clustering
} } }

\vspace{1cm}

{\rm Kikuo Harigaya\footnote[1]{e-mail address: harigaya@etl.go.jp,
e9118@jpnaist.bitnet}}\\

\vspace{1cm}

{\sl Fundamental Physics Section, Physical Science Division,\\
Electrotechnical Laboratory,\\
Umezono 1-1-4, Tsukuba, Ibaraki 305, Japan}


(Received March 5 1992)
\end{center}


\Roman{table}

\vspace{1cm}

\noindent
{\bf Abstract}

\noindent
We propose a Su-Schrieffer-Heeger type electron-phonon model for $\soc$
with O defects and solve by the adiabatic approximation.  Two new
properties are obtained. (1) The dimerization becomes weaker around the oxygen.
Two localized states appear deep in the gap.  Optical transition between
them is allowed.  This accords with the recent optical absorption data.
(2) Oxygens are predicted to cluster on the surface of $\soc$.

{}~


\pagebreak

\section{Introduction}

The buckminsterfullerenes C$_N$ have been intensively studied.
They have hollow structures and their surface is composed of a
closed network of carbons.  The $\soc$ molecule has a
shape like a soccerball.  There are twelve pentagons and twenty
hexagons on the surface.  The molecule has the icosahedral symmetry.

There is a very sharp peak due to $\soc$ with a large amount
in the mass spectra data of the laser-ablated graphite [1].
However, the mass spectra of the sample in solvents indicate the
presence of oxidized molecules: C$_{60}$O$_n$ ($1 \leq n \leq 5$) [2].
The amounts of C$_{60}$O$_n$ in solvents (with and without ultraviolet
light) become larger as the exposure time is longer.
They become as large as $1-10$\% or more under the ultraviolet irradiation.
The molecular structures of the oxidized $\soc$ have been reported in
the recent experiments [3].  It has been found
that an oxygen atom makes two bonds with carbons
between which there is a short bond.  A recent calculation by the
tight-binding molecular dynamics [4] has suggested the stability of
this structure.  In other words, it is energetically
more favorable to attach an oxygen along the short (double) bond than
along the long (single) bond.

What changes take place in lattice and electronic structures when oxygen
atoms make bonds with $\soc$?  This problem is important
because it has been shown [2,3] that significant amounts of
C$_{60}$-oxides may be present under some experimental conditions.
  It would be necessary to know what will
be observed due to the presence of the oxygen defects.  The knowledge is
lacking now.  We should distinguish effects due to the
oxygen defects in the experimental data.

In this paper, we study a variant of the Su-Schrieffer-Heeger
(SSH) model [5] which has been used in studies of conjugated polymers.
The application of the model for $\soc$ with no defect has been successful
[6,7].  First, we consider a geometry shown schematically in Fig. 1.  An
oxygen atom is attached to two sites where there would be a short bond
otherwise.   We show a large intrusion of two localized states into the gap.
This change will give rise to new observations.
Symmetries of wave functions indicate that optical transition between them
is allowed.  Next, geometric configuration of many oxygens on the surface
is investigated.  A quite new property, i.e., oxygen clustering is
certainly predicted for all the parameters we take.  Consequences for
experiments are discussed.

\section{Model and formalism}

The SSH model [5] is extended to $\soc$ with an O defect,
\beeqa
H &=& H_{\rm C60} + H_{\rm O},\\
H_{\rm C60} &=& {\sum_{\langle i, j \rangle,s}}^{'}
[- t + \alpha (u_i^{(j)} + u_j^{(i)})]
(c_{i,s}^\dagger c_{j,s} + {\rm h.c.}) \nonumber \\
&+& \frac{K}{2} {\sum_{\langle i, j \rangle}}^{'}
(u_i^{(j)} + u_j^{(i)} - C )^2,\\
H_{\rm O} &=& E_{\rm O} \sum_s d_s^\dagger d_s
- t_{\rm O} \sum_s [ d_s^\dagger (c_{1,s} + c_{2,s}) + {\rm h.c.} ].
\eneqa
In the first term of $H_{\rm C60}$,
the quantity $t$ is the hopping integral of the
system with the uniform bond length;
$\alpha$ is the electron-phonon coupling; the
operator $c_{i,s}$ annihilates a $\pi$-electron at the $i$-th carbon atom
with the spin $s$; particularly the labels $i = 1, 2$ are used for two carbons
adjacent to the oxygen atom, as shown in Fig. 1; $u_i^{(j)}$ is the
displacement of the $i$-th atom in the direction opposite to the $j$-th atom;
the sum is taken over nearest neighbor
pairs of $\langle i, j \rangle$.  The quantity, $u_i^{(j)} + u_j^{(i)}$,
is the change of length of the bond between the $i$- and $j$-th atoms.
When it is positive, the bond becomes longer and
the hopping integral decreases from $t$;
accordingly we take the sign before $\alpha$ to be positive.
The prime symbol at the summation indicates that the pair of nearest neighbor
atoms adjacent to O, $\langle 1, 2 \rangle $, is not included in the sum.
The second term is the elastic energy of the
phonon system; the quantity $K$ is the spring constant.
We include the constant $C$ in order to remove the contraction of the
lattice; it is determined to reproduce the previous calculation [6] of the
undoped $\soc$.  The term $H_{\rm O}$
represents the interaction between the $\soc$
molecule and the oxygen.  As the details of characters of the bonds
between the oxygen and carbons are not well known, we assume
one effective orbital which forms two bonds with $\pi$-electrons of carbons.
This is a valid assumption because the atomic levels of the oxygen
would be present deep in the occupied states of $\soc$, and
the electronic structures around the energy gap of $\soc$ would not
sensitive to details of the electronic structures of the oxygen.
The electronic states at
the oxygen are denoted by the operator $d_s$ with the atomic energy
$E_{\rm O}$.  The quantity $t_{\rm O}$ is the hopping integral between
the carbon and oxygen atoms.  We assume the electronic states
with the nearly-closed shell.  This is parametrized by the energy $E_{\rm O}$
which is much deep from the Fermi energy.  The $d$-states would be filled with
nearly two electrons.  We do not include an intra-site Coulomb interaction
because we limit our discussion to static electronic structures.
This assumption would be more or less valid when there are even number
of electrons.  The interaction should be taken into account when we investigate
dynamical properties in detail.

The model eq. (1) is solved by the adiabatic approximation for the lattice.
The eigenvalue equation for electrons is
\beeqa
\eps_\kappa \phi_{\kappa,s} (i) &=& \sum_{\langle i, j \rangle}
(-t + \alpha y_{i,j} ) \phi_{\kappa,s} (j) \mbox{~~~~for $i\neq 1,2$},\\
\eps_\kappa \phi_{\kappa,s} (i) &=& {\sum_{\langle i, j \rangle}}^{'}
(-t + \alpha y_{i,j} ) \phi_{\kappa,s} (j) - t_{\rm O} \chi_{\kappa,s}
\mbox{~~~~for $i=1,2$},\\
\eps_\kappa \chi_{\kappa,s} &=& E_{\rm O} \chi_{\kappa,s}
- t_{\rm O} [ \phi_{\kappa,s} (1) + \phi_{\kappa,s} (2) ],
\eneqa
where $\eps_\kappa$ is the eigenvalue of the $\kappa$-th eigenstate,
$\chi_{\kappa,s}$ is the amplitude at the O site,
and $y_{i,j} = u_i^{(j)} + u_j^{(i)}$ is the bond variable.
The self-consistency equation for the lattice is
\beeq
y_{i,j} = - \frac{2\alpha}{K} {\sum_{\kappa,s}}^{\rm occ} \phi_{\kappa,s}(i)
\phi_{\kappa,s}(j) + C,
\eneq
where the notation, occ, means the sum over the occupied states.
Equations (4--7) are solved by the numerical iteration method used in the
previous publications [6].

\section{Results}

All the quantities with the energy unit will be shown in the unit of $t$.
We take $\alpha/t = 2.52$\AA$^{-1}$ and $K/t =19.9$\AA$^{-2}$; these give
the dimensionless electron-phonon coupling: $\lambda \equiv 2\alpha^2
/ \pi K t = 0.20$ [6].  For the parameter $C$, we use $C = 0.0131164$\AA.
Number of electrons is $N_{\rm el} = 62$; the sixty electrons are
from $\soc$ and the other two are from the oxygen.
The realistic value of $t$ is about 2.5eV.  We have used it in [6].

Figure 1 shows the lattice geometry.  The $\pi$-conjugation is broken
along the bond between the sites, 1 and 2, in the present model.
The length difference between the
short and long bonds of two hexagons adjacent to these sites becomes smaller.
We cannot distinguish between the short and long bonds clearly.  Therefore,
the bonds of two hexagons are shown by dashed lines in Fig. 1.  In other
words, the dimerization becomes weaker around the O defect.
When the site is further away from the defect,
the dimerization strength becomes nearer
to that in the perfect $\soc$.  We do not show each bond length explicitly
because it is very complex to depict twenty-five different lengths.
However, it is very likely that the dimerization
is the most perfect around sites
which are located at the side opposite to the oxygen in the soccerball.

Next, we discuss localized states due to the defect.  As the phenomenological
tightbinding parameters appropriate for the defect are not known well, we
vary them arbitrary.  The main conclusion---the presence of the
localized states---does not change for all the parameter ranges we take.
We report results for the parameters: $t_{\rm O} = 0.5t, 0.75t$ and
$-2t \leq E_{\rm O} \leq 0$.  Figures 2(a) and (b) are for $t_{\rm O}
= 0.5t$ and $0.75t$, respectively.  In Fig. 2, the highest occupied molecular
orbital (HOMO) and the lowest unoccupied molecular orbital (LUMO) are
shown by full and open circles, respectively.  The next HOMO (NHOMO) and next
LUMO (NLUMO) are represented by the small
squares which are connected by curves.
Other energy levels, which are lower than the NHOMO or higher than the
NLUMO, are not shown.  The small circles will be explained in a later
paragraph.  When $E_{\rm O}$ is varied, the energies of NHOMO and NLUMO do
not change so much as those of HOMO and LUMO. This is the consequence of the
fact that the wave functions of NHOMO and NLUMO spread almost over
the $\soc$ while those of HOMO and LUMO are localized around the defect.
Thus, we can regard the HOMO and LUMO as impurity states which are well
known in bulk semiconductors and superconductors.  The energy difference
between the NHOMO and NLUMO, about $0.8t$, is very close to the value
of the perfect $\soc$.  Therefore, we can say that the energy gap of $\soc$
itself is less affected by the defect while two new localized states are
emitted into the gap due to the defect potential.  This property is the same
as in impurity states of bulk materials.

We look at details in the parameter dependence.  When the site energy
$E_{\rm O}$ is deep enough in the bonding states, the intrusion of the
two localized levels in the gap is small.  As $\eo$ increases, the HOMO
moves upward but the energy of the LUMO does not change so much.  The level
crossing between them occurs at $\eo \sim -0.5 t$
and $\eo \sim -1.1 t$ in Figs. 2(a) and (b), respectively.
For larger $\eo$, the energy of the HOMO is almost constant but
the LUMO moves upward.  The wave function of one of the localized levels,
which moves largely as $\eo$ changes, has a large amplitude at the oxygen site.
It is totally symmetric with respect to the oxygen.
On the other hand, the wave function of the other level has a negligible
amplitude at the oxygen.  It is totally antisymmetric.
The difference in the variation of energies of localized states, i.e.,
whether the energies depend largely on $E_{\rm O}$, is
well explained by whether the amplitude of the wave function is negligible
at the oxygen, or not.  The symmetries of the HOMO and LUMO are opposite
with each other.  This indicates that optical absorption is allowed between
two levels.  This is the consequence of the reduced symmetry from that
of the pure $\soc$.  In Fig. 1(b) of ref. [3], the optical absorption spectrum
of C$_{60}$O has been reported.  A weak peak appears at the wave length
680nm, while it is absent in the spectrum of $\soc$.  The energy corresponding
to this peak is 1.82eV.  This is close to that of the HOMO-LUMO gap
of $\soc$.  The forbidden transition between the HOMO and LUMO in $\soc$ [6]
would become allowed in C$_{60}$O.  This property accords with the above
result of the present theory.

In order to explain why two localized states appear in the gap, we consider
a $\soc$ molecule, eq. (2), with a bond impurity, $I_{\rm b} \sum_s
( c_{1,s}^\dagger c_{2,s} + {\rm h.c.} )$, between sites where the oxygen
atom exists in Fig. 1.  The impurity potential of the strength $I_{\rm b}$
modulates the hopping integral between nearest neighbor sites, and
therefore it is called the bond impurity.  The case $I_{\rm b} = - t_{\rm O}^2
/ E_{\rm O}$ is calculated.  We obtain two localized levels in the gap
again.  The results are shown by the small circles in Fig. 2.  The
tendency of the variation of energy levels is similar to that of the
previous calculation.  This is due to the fact that the model eq. (3) is
transformed into the bond impurity with $I_{\rm b} = - t_{\rm O}^2
/ E_{\rm O}$ when the second order perturbation with respect to
$t_{\rm O}$ is performed.  The perturbation becomes better when the
site energy $E_{\rm O}$ is much deeper.  In fact, the separation between
small and large circles becomes smaller for deeper $E_{\rm O}$ in Fig. 2.
Why do two energy levels appear in the gap?
The positive impurity strength $I_{\rm b}$ corresponds to a negative
hopping integral between sites, 1 and 2.  Due to this defect potential,
two localized levels are emitted in the gap; one is from the bonding
states and the other from the antibonding states.  Wave functions of these
levels are localized around the bond impurity.  The similar property
has been discussed in the work of impurity effects in conjugated
polymers [8].

Finally, we report many oxygen cases.  We first carry
out the calculation for two
cases: one case is that two oxygens are attached along short bonds, and
the other case is that one is attached along a short bond while the other along
a long bond.  Mutual positions of oxygens are changed, and all the positions
of the entire $\soc$ molecule is taken into account.
In both cases, it is energetically most favorable that the two oxygens are
connected along sides of a common hexagon.   In the latter case,
two oxygens are connected along nearest-neighbor sides of a hexagon.
The energy of this solution is lower than that of the optimum solution of
the first case.  Thus, the oxygens tend to cluster on
the surface of the $\soc$ molecule.  The reason is that the mixing of
energy levels, which are localized around the two oxygens,
becomes larger as the distance
between them is shorter.  It is energetically  most favorable to maximize
the bonding-antibonding splitting of energy levels.  The energy gain
from the solution with the next lowest energy is about $0.5-1$eV.
In the calculation, the oxygen clustering
persists for all the values of $\eo$ we take.
Second, the calculation is extended to systems with three oxygens.
We always obtain clustered configurations of oxygens in this case.
Therefore, the clustering would be observed in experiments.

\section{Concluding remarks}

In the present paper, we have not included a hopping integral $t_{1,2}$
between sites, 1 and 2.  We have assumed covalent bondings between the
oxygen and neighboring carbons.  Even if it will exist, the magnitude
of $t_{1,2}$ will be smaller than that of $t$. In $\soc$ without the
oxygen, there is a short band between the sites, 1 and 2.  The larger
hopping integral is assigned to this bond.  The hopping $t_{1,2}$
(which are smaller than that of the short bond) in C$_{60}$O is
effective for reducing the local gap at the bond.  This gives rise
to the intrusion of the two localized levels in the gap, also.
The situation is the same as that in the effects of the bond impurities
in conjugated polymers [8].  Thus, our conclusion
is not affected by the presence of $t_{1,2}$.

In the discussion of the impurity clustering,
we have not included direct Coulomb
interactions among oxygens and $\soc$.  The conclusion would be valid
as far as we deal with the static structures of the molecules.
The similar types of covalent bond models were used for defect states
(dopants [9], carbonyls ($>$C$=$O) [10], and $sp^3$ defects [11])
in conjugated polymers. The clustering was certainly concluded
in polymers, too [9,11].

In summary, the simple electron-phonon model has been proposed for $\soc$
with oxygen defects.  Principal conclusions are: (1) The dimerization becomes
weaker around the oxygen.  Two localized levels appear deep in the gap.
Optical transition between them is allowed. This would have been observed
in the recent experiment [3]. (2) Possible oxygen arrangements have been
studied by minimizing the energy.  Oxygens tend to cluster for all the
assumed parameters.

\noindent
{\bf Acknowledgements}\\
Fruitful discussion with Dr. K. Yamaji, Dr. S. Abe, and Dr. Y. Asai,
is acknowledged.   The helpful information from Dr. Y. Tanaka is also
acknowledged. Numerical calculations have been performed on FACOM M-780/20
of the Research Information Processing System, Agency of Industrial
Science and Technology, Japan.

\pagebreak

\begin{flushleft}
{\bf References}
\end{flushleft}

\noindent
$[1]$ Kroto H W, Heath J R, Brien S C, Curl R F, and Smalley R E 1985
{\sl Nature} {\bf 318} 162\\
$[2]$ Wood J M, Kahr B, Hoke II S H, Dejarme L, Cooks R G,
and Ben-Amotz D 1991 {\sl J. Am. Chem. Soc.} {\bf 113} 5907\\
$[3]$ Creegan K M, Robbins J L, Robbins W K, Millar J M,
Sherwood R D, Tindall P J, and Cox D M 1992 {\sl J. Am. Chem. Soc.}
{\bf 114} 1103\\
$[4]$ Menon M and Subbaswamy K R 1991 {\sl Phys. Rev. Lett.} {\bf 67} 3487\\
$[5]$ Su W-P, Schrieffer J R, and Heeger A J 1980 {\sl Phys. Rev. B} {\bf 22}
2099\\
$[6]$ Harigaya K 1991 {\sl J. Phys. Soc. Jpn.} {\bf 60} 4001;
{\sl Phys. Rev. B} {\bf 45} (to be published in June 15 issue)\\
$[7]$ Friedman B 1992 {\sl Phys. Rev. B} {\bf 45} 1454\\
$[8]$ Harigaya K, Terai A, Wada Y, and Fesser K 1991
{\sl Phys. Rev. B} {\bf 43} 4141\\
$[9]$ Harigaya K 1991 {\sl J. Phys.: Condens. Matter} {\bf 3} 8855\\
$[10]$ F\"{o}rner W, Seel M, and Ladik J 1986 {\sl Solid State Commun.}
{\bf 57} 463; Harigaya K 1991 {\sl J. Phys.: Condens. Matter} {\bf 3}
4841\\
$[11]$ Jeyadev S and Conwell E M 1988 {\sl Phys. Rev. B} {\bf 37} 8262\\

\pagebreak

\begin{flushleft}
{\bf Figure Captions}
\end{flushleft}

\noindent
Fig. 1. $\soc$ with an O defect.  The numbers, 1 and 2, indicate two carbons
making bonds with the oxygen.

{}~

\noindent
Fig. 2. Energy level structures in the gap as a function of $E_{\rm O}$.
We use $t_{\rm O} = 0.5t$ in (a) and $t_{\rm O} = 0.75t$ in (b).
The HOMO and LUMO are shown by full and open circles, respectively.
The NHOMO and NLUMO are represented by small squares.  The small circles
are the HOMO and LUMO of $\soc$ with the bond impurity of the strength,
$I_{\rm b} = -t_{\rm O}^2 / E_{\rm O}$.

{}~

\noindent
Please request figures via e-mail.  They will be sent via air-mail.

\end{document}